# Government mandated blocking of foreign Web content[*]

Maximillian Dornseif
md@hudora.de - http://md.hudora.de/

**Abstract:** Blocking of foreign Web content by Internet access providers has been a hot topic for the last 18 months in Germany. Since fall 2001 the state of North-Rhine-Westphalia very actively tried to mandate such blocking. This paper will take a technical view on the problems imposed by the blocking orders and blocking content at access or network provider level in general. It will also give some empirical data on the effects of the blocking orders to help in the legal assessment of the orders.

## 1 The Problem

The blocking order was decreed by the district government of Düsseldorf and is trying to force access providers to block US-hosted Web pages, it stirred considerable discussion in the German legal and civil rights communities. The orders raise considerable legal and policy issues which are being discussed broadly in literature and in front of several courts.

Being a "technopolicy" issue, the blocking orders involve a complex interplay between engineering, applied psychology, law and economics. While the legal issues have been discussed in considerable detail by scholars, administration, civil right groups, lawyers and courts there has been no serious discussion about other aspects of the issues involved in the blocking order.

This paper will try to shed some light on the engineering issues involved. Network engineering issues of the blocking orders have seen no serious analysis up to now. Legal literature mainly refers to other legal literature for these issues, which has resulted in soma lack of technical depth in the discussion.

Little is known about the workings and side-effects of the different techniques discussed for blocking. While IP-filtering is at least well studied in the context of network security and considerable commercial experience exists with filtering HTTP proxy software, using recursive DNS servers to fake data in zones delegated to other entities is an obscure technique which hasn't seen any serious research so far.

To analyze the network engineering issues of the blocking orders it will be first looked into different techniques which could be used to deploy blocking at provider level, including

---

their actual workings, pitfalls and circumvention measures. Then we will try to find what the blocking order actually asks for. Finally we will survey if and how providers comply with the order.

## 2  Techniques of blocking

To take a look at techniques which can be used for blocking it must first be determined what is meant by "blocking". Blocking should be defined in scope of this work as refusing users access to certain web pages without the cooperation of the content provider, the hosting provider and the the owner of the client machine being used to access these pages. This will set a focus on measures implemented at the access provider or backbone provider level. Both kinds of companies will be summarized as providers in the rest of this paper.

Filtering measures at the providers can be categorized in three groups working at two different OSI-Model layers: IP-filtering attacks at the networking layer to deny traffic with the blocked hosts. DNS-tampering doesn't block the inter-host communication itself, but keeps the client from finding the information needed to initiate communication with the blocked host name. Filtering Proxies are interposed in the communication and a wide array of modifications to traffic passing them including the denial of access to blocked content completely can be accomplished with this technique.

It will be now looked into the technical details of each of these ways of blocking including their implementation, cost, granularity, side effects and circumvention.

### 2.1  Packet Filtering

Packet filtering describes a technique used by routers and other devices embedded in the data stream deciding whether to allow or deny further passage to single packets by looking at the headers of the packets while they pass through them. Two types of packet filtering relevant to blocking of Web hosts can be distinguished: layer 3 filtering and layer 4 filtering. While generally based on the same principles the two methods differ in their granularity and resource consumption on the filtering device.

Packet filtering functionality is generally already available in routers deployed between provider's networks to provide mechanisms for security precautions like egress and ingress filtering. This means that providers in principle do not need to buy new hardware to deploy packet filtering based blocking.

Packet filtering offers no way of informing users trying to access blocked content of the blocking measures and, even more important, there is no way of informing users trying to access services which were not intended to be blocked but actually are blocked because of coarse granularity in IP-filtering.

### 2.1.1 Layer 3 Filtering

Layer 3 filtering looks at the so called IP-header which mainly contains information about the machine sending the packet and the machine where the packet is designated. Both are encoded as so called IP-addresses. Layer 3 filters basically allows to define filtering rules which match on destination address or source address of the communication.

To block the access to Web content on a certain host, all traffic outbound to the host's address or inbound coming from the host's address can be denied. Both variants result in total suppression of all successful Web communication with the host. The reason is that already the the TCP handshake which initializes the Web HTTP connection can't be successfully completed. When only incoming traffic from the blocked host is denied, the blocked host could use special software to monitor the arriving SYN-packets as the first step in a TCP-handshake. This would give the ability to count the number of requests tried but suppressed by the blocking measure. To the user trying to access the blocked Web content, IP-filtering only in incoming direction would result in an extended wait when trying to access the blocked host followed by an error message like "Connection timed out". When traffic going out to the blocked host is denied, the owner of the blocked host has no ability to monitor connection attempts. To the user accessing the host, this would depending on the actual implementation of the filter rule mean immediately a "Couldn't connect" or a "Connection timed out" error message.

Layer 3 filtering denies practically all communication between the blocked host and the blocking providers network. This means that not only accessing Web pages is suppressed, but also all other services provided by the host, be it email, chat, Usenet or anything else. It also means that the blocked host can't request Web pages from the blocking providers network or initiate other forms of Internet based direct communication with the blocking providers network.

Layer 3 filtering uses relatively little resources on the networking devices preforming it. Routers generally have to parse addresses in the IP-header to decide further routing of packets. Nevertheless, routers can only process a fixed amount of rules effectively. Also routers in the cashed-stripped telecommunications business might already be used at their maximum capacity and can't handle any additional resource overhead at all.

### 2.1.2 Layer 4 Filtering

Layer 4 filtering uses all the information available to layer 3 filtering but also inspects the headers of the data inside the IP packets. The main information gained by this inspection are the so called port numbers.

Port numbers allow Internet hosts to offer several services using the same IP-address. Every well known Internet service has a port number assigned which is used to direct incoming data to the appropriate server software on the host. Often used ports include 21 for file transfer (FTP), 25 for sending email (SMTP), 53 for DNS, 80 for the Web (HTTP) and 110 for reading email (POP3). Web access is insofar special, because besides the standard port 80 some other ports are also used regularly for Web purposes. Port 443 is used for secure Web connections (HTTPS) and people sometimes run their Web server software on ports 8000 and 8080, which seems to be done mostly for Unix implementation specific reasons.

By basing the filtering decision not only on the destination IP-address but also on the destination port number, it is possible to do blocking on a per service basis. This means while blocking all direct Web access of content on the blocked host, users still would be able to use other services on it, like email and chat. Blocking on layer 4 will also practically not affect attempts from the blocked host to access services in the blocking provider's network.

User experience of layer 4 filtering when accessing blocked contents is generally comparable with the effects of layer 3 filtering described above.

Network devices do not have to parse layer 4 headers for their routing decision. This results in a theoretical increase of processing needed when activating layer 4 filtering.

While layer 4 filtering can be used to focus blocking only on Web Access by just denying traffic to port 80 and possibly 443, 8000 and 8080, it still works on an all-or-nothing basis with regard to which Web pages will be blocked. There is no control at all on the degree to which content on the host using the blocked IP will be suppressed. Accessing the volume of content unintentionally blocked together with the content blocked intentionally is a difficult problem.

An obvious problem is content by the same content provider presented in the same publishing context. We will call this a "Website". Usually only certain pages of a site will give reason for blocking, but IP-filtering will block all content on that Website. In the last century it was very likely that a Web host contained a large number of sites in different subdirectories. While using a different domain name but the same IP-address for every Website at a hosting provider has become widespread nowadays, we still see Web hosts serving dozens and sometimes thousands of Websites using the same domain name. A prime example are Universities' Web servers, where each faculty usually runs its own Web server on which each institute presents its information in a subdirectory upon its own responsibility as an independent Website. An other famous example is www.xs4all.nl which hosted between 3000 [Pro97] and 6000 [Sie99] different Websites when it was blocked in 1996/1997 by German providers because of about twp dozen illegal pages.

Another technique for hosting several sites on a single host is so called "name based virtual hosting" which is mandatory supported by the actual standard for Web access named HTTP/1.1 [FGM$^+$99], but was also widely supported in HTTP/1.0 compliant software [APN00]. With name based virtual hosting, a host can provide content for several different Web sites using different domain names but sharing a single IP-address.

This means that blocking a single IP-address can deny access to several domain names at once. This problem gets worse by the fact that there is without the cooperation of the hosting provider no reliable way of determining how many domain names and which domain names are hosted at a certain IP-address. A recent study by Benjamin Edelman found that more than 87% of all domains deploy name based virtual hosting for their primary Web server [Ede03]. Edelman surveyed only the Top-Level Domains com, net and org and only searched for the most common host name www. This indicates that the actual number of sites sharing IP-addresses is even higher. In [Ros03b] it is reported that www.nazi-lauck-nsdapao.org is served by virtual hosting. My own research shows that at least parts of the stormfront.org domain deploy virtual hosting.

## 2.2 DNS-Tampering

Tampering with the Domain Name System (DNS), a worldwide distributed database now seems to be the preferred way of blocking - at least the district government and the courts focus on this in their legal arguments. The whole area of DNS is plagued by inconsistent use of terminology. So we first look into the terminology and workings of DNS, then at the ways of deploying blocking and the problems in doing so.

In Internet based communication, computers are addressed by their IP-address which is uniquely identifying a host taking part in Internet communication. The IP-address is a number roughly in the range between 0 and 4 billion, usually written as four numbers separated by dots, e.g. 172.19.234.11. Since these numbers seemed inappropriate for human use the DNS was introduced as a distributed telephone book like service mapping names to addresses. DNS database lookups are nowadays so tightly integrated into applications and operating systems that users can be completely ignorant of IP-addresses and use only domain names. Even programming languages and APIs like the Python socket module start hiding the existence of IP-addresses from programmers.

Still every use of Internet communication using a domain name first results in a DNS lookup and after that in the requested connection itself to the resulting IP-address as shown in Figure 1. DNS-tampering does not address the denial of transport to the actual data from the blocked host to the user requesting the content.

DNS-tampering on todays Internet can't distinguish between different services although wider deployment of so called DNS SRV resource records might change that. The actual situation means that many effects of DNS-tampering are similar to layer 3 IP-filtering: All services of the blocked domain name, be it Web, chat or file transfer, will be inaccessible. The only exception is email which will be routed not only via the so called DNS A resource records used for other services, but also via DNS MX resource records used exclusively for email.

### 2.2.1 Terminology

First of all: up to now nobody has advocated the blocking or filtering of DNS traffic. All proposals center around answering DNS queries directed to the providers recursive DNS servers with replies not representing the actual data saved in the distributed DNS database. This might be viewed as corruption of requests to the distributed DNS database or as faking DNS replies. Thus the term "DNS-tampering" is more appropriate than "DNS-blocking".

An other source of confusion is the usage of the term "DNS server". This term is often used for software preforming two completely different functions. The fact that the popular BIND software package integrates this two functions in a single piece of software and thus both functions are sometimes preformed by the same host further fosters confusion. Even best-of-breed technical literature on DNS like [AL99] uses imprecise language on this matter.

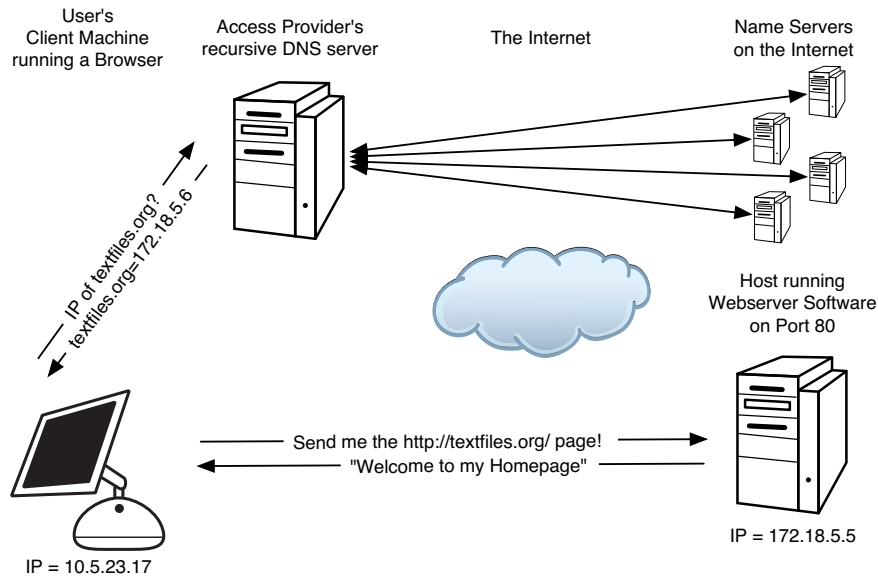

Figure 1: Working of recursive DNS servers

To reduce confusion, naming convention based on [Moc87] Section 2.2, the standard defining DNS, should be used:

**Name Server** The Purpose of name servers is the publishing of DNS data. They are part of the distributed DNS database and usually maintained by hosting providers or their associates. Persons not involved in the management of networks usually never get in direct contact with name servers.

**Recursive Server** Recursive Servers, also known as "recursive resolvers" query name servers on behalf of their clients. The conversion of an name to an IP-address is usually done by querying name servers in a recursive fashion. This means that the first name server queried (the so called "root name server") returns a reference to another name server which can provide further detail and will provide reference again until ultimately a name server is found which can provide the actual data being searched for. The piece of software carrying out the query process is called a "resolver". It can be seen as a database query tool querying the distributed DNS database. In instances where a client machine does not implement a resolver on its own, it contains a "stub resolver" which sends queries just to a recursive server which will carry out the query process on the clients behalf and return the final result to the client. See figure 1 for an overview.

Technically speaking recursive servers can be seen as DNS proxy servers. An end user may come in contact with recursive servers when configuring his Internet access. Recursive servers are often called "DNS-Servers" in general literature or in Operating System configuration dialogs. To accommodate this recursive servers will be called "recursive DNS servers" in the rest of this paper.

Further confusion arises from the fuzzy terminology relating to domain names. The terms "host name", "host", "domain name", "domain" and "site" are often used as if they were interchangeable. Following definitions are based on [Moc87] and [AL99].

**Domain Name**  A fully qualified domain name (FQDN) is a dotted string unambiguously pointing to an object in the global DNS database. For brevity it will be assumed in this paper that all domain names are FQDNs.

**Domain**  A domain is a subtree in the distributed DNS database. It can be identified by a domain name. All domain names ending in this domain name are part of the domain.

**Zone**  A zone is a possibly pruned subtree in the distributed DNS database. Zones contain at least one domain and might or might not contain subdomains branching off that domain. For many name server and recursive DNS server software a zone is the smallest unit to which access controls, authority and the like can be applied.

**DNS Resource Record**  A domain name usually points to a handful different types of data in the distributed DNS database. The individual records of data are called DNS resource records.

**DNS A Resource Record**  DNS A resource records pointing to a host's IP-address allowing the fundamental application of the DNS by mapping names to addresses.

**DNS MX Resource Record**  DNS MX resource records are used for deciding which host will receive the email for a domain name. DNS MX resource records point to a domain name while DNS A resource records point to an IP-address.

**Host**  A host is a machine accessible from the Internet via at least one IP-address. It usually also has at least one domain name pointing to it's IP-address.

Interaction between email delivery and DNS-tampering can be extremely complex. Email delivery usually involves at least two computers running so called Mail Transport Agent (MTA) Software. The first MTA often runs on a different host than the one the user is utilizing to compose and send the email. The host running the first MTA may use a different recursive DNS server which might be tampered with differently than the one the user is utilizing. Also most MTAs under certain conditions try to use DNS A resource records instead of DNS MX resource records. To keep discussion of the effects of DNS-tampering to email to a tolerable level, in this paper only a simple case will be considered where the first MTA uses the same recursive server as the user and only deploys DNS MX resource record lookups to find the destination server.

### 2.2.2 Blocking Techniques using DNS

The DNS Protocol as defined by [Moc87] reserves a response code named REFUSED to indicate that "the name server refuses to perform the specified operation for policy reasons." While [Moc87] speaks of name servers, there is no indication that this response code shouldn't be used by recursive DNS servers.

Returning the REFUSED response code to DNS requests for the domain names of sites to be blocked is the obvious way of blocking via DNS. Users trying to access the site would see an error message along the lines of "Host not found". Configuration for doing so with the popular BIND software package is shown in figure 2. Note that you can't use this configuration technique with BIND to only refuse queries to DNS A resource records. Instead queries all other resource records in the zone. including DNS MX resource records, are refused.

```
//REFUSED configuration
zone "REFUSED.example.com" {
   type master; file "dummy.zone";
   allow-query {none;};
};

// silence configuration
zone "silence.example.com" {
    //  no name server listening on 127.0.0.1
    type forward; forwarders {127.0.0.1;};
    forward only; };

// SERVERFAIL configuration
zone "SERVERFAIL.example.com" {
    //  no name server listening on 127.0.0.1
    type slave; masters {127.0.0.1;};
    file "/tmp/dummy.zone"; };
```

Figure 2: Configuration Examples for BIND.

Other manipulations at the providers recursive server are possible and - as will be shown by empirical results - common. There are five classes of forgery which can be used to suppress communication:

**NXDOMAIN** This form of forgery returns a "Name Error", commonly called NXDOMAIN, meaning that the name which the user is asking for does not exist. Per [Moc87] this answer may only be sent by "authoritative name server" for the requested domain which the provider's recursive DNS server is not, so this is forgery of DNS data and a breach of the DNS standard. This manipulation will result in a "Host not found" error message to the user. Using BIND this behaviour is configured by creating an empty zone file and reconfiguring BIND not to run just as an recursive DNS server, but also as an authorative name server for a zone with the same name as the the domain name to block. This technique can be used in principle to manipulate single DNS resource records. Name hijacking

**Name Hijacking** Name hijacking means that recursive DNS servers do not return the data saved in the distributed DNS database but some made up data aimed at directing the client at another host than the one it was looking for. As with NXDOMAIN replies this forgery is not standard compliant. Usually the user will see a different Web page than the one he intended to access. This is configured in BIND basically the same way as NXDOMAIN replies, but the zone file is not empty but contains DNS resource records pointing to the server the hijacking should lead to. Name hijacking can be used in principle to manipulate single DNS resource records.

**Name Astrayment** Technically name astrayment is a variation of Name Hijacking but the recursive DNS server is returning an address which is deemed to be unused or illegal. This will result in the client software trying unsuccessfully to connect and return an error message like "Could not connect". Configuration in BIND is exactly the same as with name hijacking. Name Astrayment can be used in principle to manipulate single DNS resource records.

**Silence** The single way of not returning incorrect data for a blocked domain name besides returning REFUSED answers is not answer at all. Using BIND silence shares the problem of tampering with all DNS resource record types, like sending refused replies does. The Silence method of tampering will result in a "Host not found" message to the user after a considerable delay. See figure 2 for a possible implementation of this technique with BIND. It should be explicitly stated that BIND was not designed to be used this way so that this configuration can result in malfunctions of the software.

**Provoked Server Failures** Provoked server failures are from the servers point of view similar to silence. Instead of provoking a internal failure leading to a timeout, an internal failure which is detected immediately is provoked and leads to a SERVER-FAIL response to the client. See figure 2 for a possible implementation with BIND. It should again be explicitly stated that BIND was not designed to be used this way, so that this configuration can result in malfunctions of the software.

### 2.2.3 Problems of DNS-Tampering

Real-world implementations of DNS-tampering face serious problems. DNS was designed to control data on a per zone basis. Domains are delegated from domains higher in the tree of the distributed DNS database to the respective owners of the domains which manage their domains in one or more zones. Therefore, neither DNS nor popular server software for DNS are designed to allow modifications to single DNS resource records on their own, but only to DNS resource records as part of a zone you are authoritative for, meaning you "own" the zone's domain. This actually means a provider trying to implement DNS based blocking can not configure it's software to block just certain names. It must create zones on it's name servers for the names to block and put the fake data in there. It must also retrieve all other data not to be blocked from the authorative, name servers and put it in there. It finally must assure in some way that this data retrieved from the authorative nameservers is in sync with their sources in some way.

An example will illustrate the problems: providers are ordered to block all Web pages on the host names www.bad.example.com and bad.example.com. There is another host named not-so.bad.example.com which is not to be blocked. There are also several dozen hosts in the domain to.bad.example.com. Using the name hijacking technique providers could create a zone named bad.example.com and insert fake DNS A resource records for bad.example.com and www.bad.example.com in the zone file. Now they have to retrieve MX and all other kinds of DNS resource records for those two entries from the authorative name server and also add them to their zone file. Failing to copy this records from the authorative name servers would result in failures for other queries than to DNS A resource records. This would be especially harmful to email traffic and which was - contrary to Web traffic - not to be blocked.

They also have to retrieve all resource records for not-so.bad.example.com from the authorative name server and insert them in their zone file. They can get around putting the data for all the names in the to.bad.example.com domain into their zone file using special DNS NS resource records to point back to the authorative servers.

They have to ensure that the data they retrieve from the authorative name servers stays fresh and changes made at the authorative name servers propagate in their zone file. Literature suggests using intervals between 3 and 24 hours for checking the if data in the zone file needs a refresh [AL99].

Even this complex setup can't be implemented with all domains. To collect the data for the zone file one must know the names in the original domain. There are two reliable way of finding the all names in a zone. The first, a so called "zone transfer" can be requested by a client from an authorative name server and will result in a complete copy of all data in the zone. But nowadays it is considered good security practice to disallow zone transfers from unknown parties [AL99]. An alternative is the technique of walking DNSSEC resource records. But since DNSSEC is not widely deployed this technique is very seldom applicable. This means often there is no way to get the list of hostnames in a domain.

The only solution to this problems seems to develop software specially designed to do DNS-tampering. The alternative would be accepting that email to the domain names containing the content to be blocked can't be sent and other hosts in the zone can't be accessed by the users of the blocking provider. Even worse, one can not tell which or how many domain names are affected.

Other problems are specific to certain types of blocking. Refusing DNS lookup, NXDOMAIN replies, name astrayment, silence and provoked server failures all leave the users with error messages not showing the reason why they where unable to access the blocked Web pages. Also users not trying to access the blocked Web pages but other unintentionally blocked services will get inappropriate error messages. Name hijacking can lead to serious privacy problems since the host where names are redirected to can record Web ac-cesses and also email, chat and other services unintentionally blocked. Name astrayment can lead to ghost packets wandering the Internet or provoke other unwanted effects in the network, depending on which IP-address is used as an "illegal" address. Using the localhost address 127.0.0.1, which always points to the individual computer the user is actually operating, can result in various disturbing results, especially if the user is utilizing a HTTP proxy. Name astrayment and silence can lead to long periods of waiting before the user gets an error message. Applications may be unresponsive while waiting for DNS replies.

DNS-Tampering may also have side-effects on other parties. For example servers at the provider may check incoming connections on consistent DNS data. This might result in certain connections, e.g. email from the hosts with the blocked names to the blocking providers servers being denied.

## 2.3 Filtering Proxies

The third technique widely discussed for blocking content are HTTP proxies deploying filtering mechanisms. HTTP is the protocol used for transferring Web pages from the Server to the user's browser and HTTP proxies can be used to mediate this otherwise direct HTTP communication. So called caching HTTP proxies are a relatively common Internet infrastructure element, providing increased browsing speed and lowering an organizations bandwidth usage. HTTP proxies also can be used in an application level firewall providing Web access to clients who are not connected directly to the Internet.

A Web browser configured to use a HTTP proxy does not request pages directly from the Web servers the user is trying to access. Instead the browser connects, as shown in figure 3 to the HTTP proxy and asks the proxy to deliver the content the user is requesting. The HTTP proxy acquires the page on the user's behalf from the Web server and delivers it to the browser. Being directly embedded in the application layer of the communication puts a HTTP proxy in a unique position for applying filtering or blocking mechanisms. Generally Web browser software has to be modified to be able to use HTTP proxies, but all major Web browsers today support the use of HTTP proxies.

### 2.3.1 Voluntary Proxy

Voluntary proxies used as caching HTTP proxy seem to be deployed at most providers today. Nevertheless some providers use no HTTP Proxies [Hoe01]. Caching proxies are a separate part of the providers' network infrastructure and can be implemented by running proxy software on a standard workstation or by a dedicated hardware caching engine. Basically the use of these proxies is completely voluntary, but some providers use various incentives like faster access or more attractive charging for traffic via proxy to convince users to use the proxies. Preconfigured Internet access software usually is configured by the provider to use the provider's proxy servers. Auto configuration in Microsoft's Internet Explorer can be used to activate proxy usage without further user involvement.

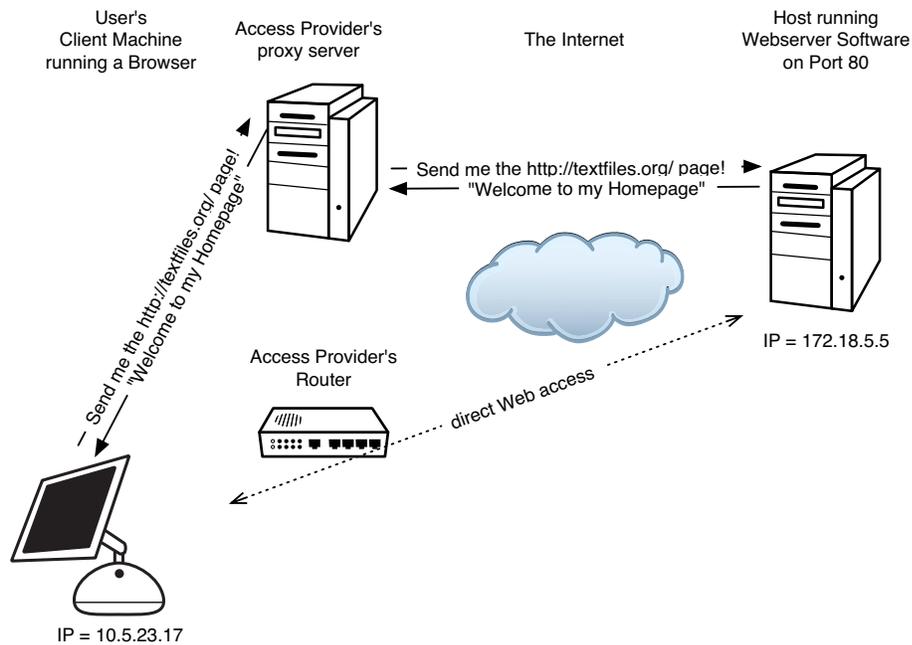

Figure 3: Working of HTTP-Proxies

### 2.3.2 Enforced Proxy

Enforced proxies are implemented exactly the same way as voluntary proxies but IP-filtering is used to deny direct Web access to users. This means users have to configure their browsers to use the provider's proxy or wouldn't be able to access the Web. To the providers this means they have the combined workload of running a voluntary proxy and IP-filtering plus considerable support costs from user inquiries.

### 2.3.3 Interception Proxy

Interception proxies - sometimes called transparent proxies - are a variation of enforced proxies. Instead of forcing the users to configure their browsers to use the proxy like in the enforced proxy model, an interception proxy architecture automatically forces every single Web related data packet through the proxy. This is implemented transparently to the user by instructing routers or other networking devices via the Web Cache Communication Protocol (WCCP) or some other mechanism to redirect Web related packets away from their destination to the proxy caches as shown in figure 4. Alternatively an in-line gateway device with cache functionality can be deployed when restructuring the network in a way that all traffic has to pass this device.

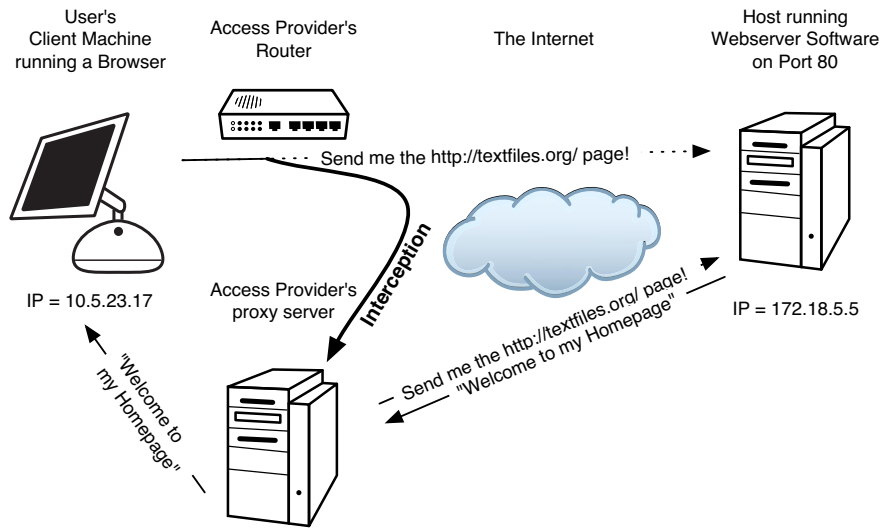

Figure 4: Working of interception HTTP proxies

### 2.3.4 Potential of HTTP Proxies

HTTP proxies substantially change the characteristics of HTTP communication. HTTP is using in principle a direct communication channel between two parties without any mediation by third parties. Using a HTTP proxy changes this by changing the direct communication to some form of mediated communication. The proxy acts as a person-in-the-middle and is able to apply any modifications to the communication and do any logging it desires. Interception proxies violate basic Internet design principles [BC01]. While his is nothing bad per se, doing so requires careful analysis is of side-effects. [Hoe01] notes that HTTP proxies increase surveillance powers and might not comply with labour protection laws.

While [Sta02b] claims they can't do, HTTP proxies are the only blocking measure which can block very exact, alt least on a per URL basis, since they have full control over the data stream. In principle HTTP proxies also allow only to filter out certain paragraphs on a page.

An remarkable experiment performed in 2000 by Dragan Espenschied and Alvar C.H. Freude helps to understand the potential of HTTP proxies [EF01]. Espenschied and Freude, at that time students at the Merz Akademie, a university of applied arts, tested the extent to which Web users are noticing manipulations done by proxies and how much they are willing to endure manipulation.

Espenschied and Freude configured most workstations in their university to use a proxy written by themselves. This special proxy was able to apply transformations to the data

stream passing through it. The transformations included:

- Exchanging words with a certain probability. This was used for mixing the names of political parties, politicians and cities, inserting links to the owners of trademarks mentioned on Web pages, exchanging "men" with "women", "China" with "Iraq", "art" with "commerce" and so on.[1]

- Extending pages of popular freemail services with elements on every page promoting the "Global Penpals Association" a fictions cooperation between Hotmail, mail.com, Yahoo! and GMX. The Global Penpals Association area appearing on every page suggested a person as personal pen pal, which the system claimed was individually "selected based on your personal options and surfing habits".

- After visiting popular search engines the user's browser was displaying an additional frame encouraging them to report anonymously found pages deemed objectionable to an anti-hate organization via an form displayed at the top of each page.

- 2% of all accesses where redirected to advertisment pages at the fake domain "InterAd.gov", which claimed to be a consortium of ICANN, Corenic, Internic, Network Solutions and the US Department of Commerce. This side was reasoning that advertisement was needed to finance the "Core-Servers" which were run by the US government. Without watching the advertisement and filling a short survey surfers were not able to reach the page they were originally trying to access.

- All students' homepages on the University's own servers appeared to have intrusive pop-up windows advertising the university.

This impressive array of modifications to users' Web reception went on mostly unnoticed by the 250 students and the university staff for several months until it was uncovered by a hardware failure in the workstation running the proxy software. Even after the experiment was publicized, most students didn't bother to change their browser's settings back so they could enjoy an unmanipulated view on the Web even through step-by-step instructions for doing so were published.

The experiment by Espenschied and Freude shows the capabilities of proxy usage for content modification. While they have not experimented with blocking pages, their research supports the assumption that substantial blocking can be done without the users noticing.

### 2.4 Circumvention of blocking measures

The legal literature discusses a wide array of circumvention techniques and often claims that circumvention is generally easy to archive [Sch99] [Zim99] [Sta03]. We will look at the different circumvention techniques mentioned and try to get to an estimate of the practicability of the various approaches.

---

[1] See http://odem.org/static/insert_coin/wordlist.txt (accessed 1.6.2003) for a complete list of exchanged words.

**Mirroring** Mirroring describes a technique commonly used to reduce resource usage on content providers' hosts and increase availability of content. A mirror is a host retrieving the content from the original source and publishing it as a Website. The retrieving and publishing is done regularly and often in an automated fashion. Mirroring can also be used to circumvent blocking by making content available from other sources than the blocked one. Mirroring is only widely done with old style static Websites which have little interactive elements. Modern, dynamic systems, e.g. so called "Web forums" or database driven websites, can be mirrored only with great effort. Mirroring can be used to circumvent IP-filtering, DNS-tampering and filtering HTTP proxies. Circumvention by mirroring is not transparent to the user: Users have to be informed of the mirrors' addresses and start using them for accessing the blocked content. Mirroring as a circumvention measure is mentioned in [Sch99], [Sie99], [Zim99] and [Eng03].

There is the assumption that users of blocked content exchange the new domain names or addresses very quickly [Sch99] [Sie99]. This needs some further research on why users which have communication channels not affected by blocking would use them to communicate new domain name instead of transporting the blocked content itself via this channels.

**Additional Domain Names** Content providers can try to evade blocking by using additional domain names pointing to the same content on the same host. While IP-filtering is immune to this circumvention technique it is effective against DNS-tampering and filtering HTTP proxies. Like mirroring this measure is not transparent to the user. Users have to be informed of the additional domain names and start using them for accessing the blocked content. In [Eng03] it is suggested that this can be done automatically by browsers but does not explain how this would work.

Additional domain names are seen relative often at blocked sites. For example http://nazi-lauck-nsdapao.com/ is also accessible via the domain names auschwitz.biz, bundesinnenministerium.biz, bundesinnenministerium.biz, bundesinnenministerium.us, bundesjustizministerium.com, bundesjustizministerium.net, bundesjustizministerium.org, bundesjustizministerium.us, bundesrepublikdeutschland.us, hitlerwasright.info, kanzleramt.us, nordrhein-westfalen.biz, nsdap.info, verfassungsschutz.biz, verfassungsschutz.us, zensurfrei.com and zuwanderungskommission.com.

**Change of IP-Address** Content providers can circumvent IP-filtering by changing the IP-address of the host running the Web server software. This should be theoretically transparent to the user, since the DNS would automatically allow the user's Web browser to start communicating with the new IP-address.

In practice implementing a way to change the IP-address several times a day needs high technical sophistication. Literature often refers to the dutch provider xs4all which in 1997 was changing IP-addresses on a hourly basis to circumvent being blocked. To asses actions taken by xs4all one must consider that xs4all is one of the oldest European providers with ties to the dutch Hack-Tic group and the German Chaos Computer Club. There would be only few places where willingness to cir-cumvent censorship and technical sophistication meet at such an high level as it did at that time at xs4all. Still negative side-effects of the frequent address changes made xs4all stop doing so after a few days. Low frequency changes of IP-address can accomplished without much technical sophistication. E.g. the host stormfront.org has changed is IP-address 6 times in the last 8 months.[2] If this happened to circumvent blocking or for other reasons is unknown.

Changing of IP-address is discussed at [Sch99], [Sie99], [Sta02a], [Eng03] and [Ros03b].

**Change of Port** Changing the port number used by the Web server software can be used to circumvent layer 4 IP-filtering and possibly filtering HTTP proxies. This can be done by simple configuration at the Web server software, but is not transparent to the user. The user has to be informed of the modified port number and start using different URLs for accessing the blocked content. E.g. http://example.com/something/ might change to http://example.com:31337/something/. This circumvention method is discussed at [Sch99], [Sie99] and [Eng03].

**Web based Services** Users can utilize translation services, privacy protection services and other Web applications which process arbitrary content referenced by user specified URL [Sie99]. Technically the services are usually a specialized kind of proxy. Using this services often requires the user to take several additional steps for every blocked page he is trying to access, so browsing a page often is impossible. Some of this services can't handle dynamic Web content or sites requiring user authentication.

**Search Engines and Archives** In the Literature there are claims that search engines could be used to access blocked content [Vas03]. It is not clear what authors are referring to when mention this. Search engines usually only show you very small excerpts of pages when displaying search results. This text fragments can't be used to circumvent filters. A single search engine - google - has an additional archive feature called "google cache" allowing the user to access a copy of the page google retrieved and saved for it's own searching purposes. This feature can be used to circumvent blocking but the cached copy of the page will be updated only every few weeks and cached copies are only saved for mostly static Web content.

The Internet Archive at http://www.archive.org/ basically offers the same functionality as the google cache, but also keeps older versions of a page instead of only keeping the most recent copy. The use of this services is not transparent to the user and requires possible special steps for each blocked page to be accessed. It should be able to circumvent all types of blocking.

**IP-Tunneling** Sometimes it is claimed that any type of blocking can be simply circumvented by using the provider to connect to an other provider and route Internet traffic through the second provider [Sie99] [Zim99]. While IP-in-IP tunneling is technically possible and can circumvent all types of blocking outlined above, there are no providers offering IP-in-IP tunneling enabled accounts to end users. Even if users would be able to find providers offering IP-in-IP tunneling on a commercial basis, software for using IP-in-IP tunneling is simply not consumer-grade nowadays. So "using a second provider" by IP-in-IP tunneling is no viable option to the average Internet user for circumventing blocking by it's provider.

---
[2]See http://uptime.netcraft.com/up/graph/?host=stormfront.org (accessed 1.6.2003).

**Proxies** Another method circumventing all types of blocking mentioned above is the use of application-level proxies - be it HTTP-, Socks- or specialized proxies like peek-a-booty or JAP or may be even alternative recursive DNS servers [Sie99] [Eng03]. These techniques require a third party cooperating in running the proxy software, which sometimes might pose a problem. After a one time installation the usage of such techniques usually is transparent to the user. Some of these like JAP seem in relative wide deployment.

**Trans-National Dial-In** Some authors discuss using long-distance or trans-national calling to access a non-blocking access provider in a different state or country [Zim99]. While doing so is possible, one must keep in mind that this is not possible for high-speed Internet access and that the cost are prohibitive to most users.

**Encryption** Several authors claim that encryption could be used to circumvent blocking [Sch99] [Eng03]. While encryption can be a powerful addition to some of the techniques mentioned above there seems to be very limited benefit for circumvention from using encryption alone. IP-filtering at layer 3 is completely unaffected while layer 4 filtering could be circumvented with certain kinds of encryption. The cryptographic DNSSEC functionality could be used to detect DNS-tampering but not to circumvent it. HTTPS could be used to hide the exact URLs being accessed from the proxy but can not be used to hide which host is accessed. All this cryptographic techniques also face non-trivial implementation issues on the severer side. Encryption without the involvement of third parties generally can hide the content of the communication but not which hosts are communication with each other, so usage for circumventing blocking measures is very limited.

**Direct Use of IP-Addresses** Literature suggests that DNS-tampering can be circumvented by using IP-addresses in URLs instead of domain names [Sie99], [Sta02b] [Eng03]. While substituting domain names with IP-addresses to circumvent DNS is generally possible, it usually can't be used for accessing domains hosted by name based virtual hosting.

There is a wide variety of possible circumvention techniques. Still it is inappropriate to state that blocking can be easily circumvented. Most circumvention techniques are technically complex or burdensome to the user or need the cooperation of a third party.

## 3 The Düsseldorf blocking orders

The actual incident driving the discussion about blocking of foreign Web content are measures taken by authorities in Germany. A short overview about the events so far will be given to allow non German speaking readers to understand the situation.

### 3.1 Prelude

The district government of Düsseldorf is the government agency for regulating media in North-Rhine-Westphalia (NRW), Germany's biggest state with about 18 million citizens.

At the beginning of October 2001 the district government invited NRW's providers for a hearing about "illegal and harmful content on the Internet". A few days later there was a second hearing on possible technical implementations to block the domains front14.org, rotten.com, nazi-lauck-nsdapao.com and stormfront.org.

At the end of November 2001 some providers started to implement DNS-tampering. Exact details of this incident are only known in regard to the City Carrier ISIS. ISIS used DNS-tampering to direct users trying to access Web content at the domains mentioned above to an Web server running at ISIS. This server used a HTTP-redirect to send surfers to a special page at the district government's Web server for "reporting right wing extremism, and other objectionable content / consumer protection". After an public outcry about the privacy implications of this, ISIS stopped the DNS-tampering. Doing so ISIS found itself accused of helping right wing extremism, so they started again with DNS-tampering. This time they made all relevant DNS entries pointing to 127.0.0.1 which can result in unexpected effects.

### 3.2 The Blocking order

In the beginning of February 2002 the district government issued a blocking order to 78 providers in NRW asking them to block two Web pages.

Following will be the parts of the blocking order which are relevant to the question what exactly is meant to be blocked as translated by a professional translator:

> 06. February 2002: Supervision under the Media Services Treaty between the German Länder [states] / Regulatory proceedings where provisions of the Media Services Treaty have been violated / Re my letter of 5 October 2001 and the oral hearings of 13 November 2001 and 19 December 2001 /Blocking order
>
> I hereby instruct you to block the access to the Internet pages [Seiten] http://www.stormfront.org and http://www.nazi-lauck-nsdapao.com as part of the services you provide.
>
> Additional note for universities and other institutions of higher education [Fachhochhochschulen]:
>
> You are exempt from this obligation to block to the extent that you need to make these sites available for the purpose of scholarship, research or teaching. If this is the case, suitable technical measures must be employed to restrict the time and location of use.
>
> Reasons: [...]
>
> Under section 18 (2) of the Media Services Treaty [MDStV] and section 1 (2) of the Regulatory Authorities Act [OBG], if there is a violation of the provisions of the Media Services Treaty [MDStV], the regulatory authority that is responsible must take the measures against the provider that are necessary to remove the violation. In particular, it may forbid providers to offer certain content and may order this content to be blocked.

The Internet pages [Seiten] listed below contain material that is prohibited by section 8 (1) of the Media Services Treaty [MDStV] and therefore violates the Media Services Treaty [MDStV]:

a) http://www.stormfront.org

The U.S. ISP stormfront.org hosts Internet pages [Seiten] that are exclusively rightwing extremist, most of them in English. A commercial service is provided in the form of various packages. stormfront.org offers storage, data transfer, email addresses with their own domain names etc., all at customary market prices.

The main page [Hauptseite] makes it clear that stormfront.org has rightwing extremist ideas. In a German language section [Angebot], for example, there is an explanation of the term "liberated zones" [Befreite Zonen] and what is intended to be done with those of a different opinion: "...we punish deviants and enemies..."

On the main page, there are links to fifteen different categories, topics and services. For example, there are a women's page and a children's page, pages on fundamental questions of racist ideology, a text library, a press page, a page for upcoming events and so on. As a whole, the site is addressed to the general public and aims to shape and influence public opinion. The site as a whole is designed, like a newspaper, with various editorial categories [redaktionell: editorially].

Because it has so many links, the site also functions as a distribution network for the German rightwing extremist scene.

This site is impermissible for the following reasons:

1. it violates provisions of the German Criminal Code [StGB] (section 8 (1) no. 1 of the Media Services Treaty)

    The pages http://stormfront.org/german/zonen.htm and the links http://www.thulenet.com/index1.php and http://www.thule-net.com/strafbar/div.htm constitute the offence of stirring up hatred against national, ethnic, racial or religious groups [Volksverhetzung] contrary to section 130 (1) and (2) of the German Criminal Code [StGB] [...]

    The pages http://www.stormfront.org/gus.html and the link http://www.panzerfaust.com/flags/flags.htm constitute the offence defined in section 86 of the German Criminal Code [StGB] (Dissemination of means of propaganda of unconstitutional organizations). [...]

2. it glorifies war (section 8 (1) no. 2 of the Media Services Treaty [MDStV]):

    for example, on the web page [Webseite] http://www.stormfront.org/german/zentrale.htm, but also throughout the whole site [Angebot].

3. in the alternative, in addition it is clearly suited to create serious moral danger to children and young people (section 8 (1) no. 3 of the Media Services Treaty [MDStV]):

    The whole Website [Angebot] of stormfront.org serves to propagate National Socialist ideology with the goal of creating National Socialist rule. [...]

[...]

b. http://www.nazi-lauck-nsdapao.com

This page makes National Socialist propaganda material available, and the victims of the holocaust are cynically disparaged. [...]

The multi-paged home page contains several pages, subdivided into categories, topics and services. The site [Angebot] as a whole is designed editorially [redaktionell], like a newspaper.

Because it has so many references [Verweise] and links, the site also functions as a distribution network, inter alia for the German rightwing extremist scene.

This site is impermissible for the following reasons:

1. it violates provisions of the German Criminal Code (section 8 (1) no. 1 of the Media Services Treaty [MDStV])

    The home page itself and the direct references [Verweise] constitute the offence of stirring up hatred against national, ethnic, racial or religious groups contrary to section 130 (1) and (2) of the German Criminal Code [StGB] by inciting hatred and destruction of Jews and other "enemies of the people". [...]

2. it glorifies war (section 8 (1) no. 2 of the Media Services Treaty [MDStV]) [...]

3. in the alternative, in addition it is clearly suited to create serious moral danger to children and young people (section 8 (1) no. 3 of the Media Services Treaty [MDStV]):

    The whole Website [Angebot] of nazi-lauck-nsdapao.com serves to propagate National Socialist ideology with the goal of creating National Socialist rule. [...]

[...]

On the basis of the technical research I have carried out and the hearing procedure, I have come to the conclusion that in the present state of the art there are three possibilities of blocking this material:

1. Excluding domains on the domain server (DNS)

    If the access provider operates a DNS, this can be configured to relay enquiries not to the correct server but to an invalid or to another predefined page [Seite].

2. Use of a proxy server

    The URL, the precise allocation criterium of the individual web page [WebSeite] on the relevant server, can be blocked by the use of a proxy server. Enquiries for the prohibited sites [Angebot] are filtered and access is denied, or the enquiry is rerouted to a predefined page [Seite] in the browser and linked [hingewiesen]:

3. Excluding IPs by blocking them at the router level

    The router can be configured so that all the data traffic to a particular IP-address is not forwarded.

In addition, further technical possibilities are being tested at present.

The three possibilities of blocking mentioned above have the effect, without making technical alterations to a user's computer, that when the above domain names, which are the subject of this blocking order, are entered, the prohibited sites can no longer be accessed. Apart from this, when notification was made that the sites were prohibited, some providers blocked the sites, most of them using the DNS method, and this showed that it is technically possible to block the sites.

The blocking is also reasonable.

It is necessary here to weigh the burden caused to a provider by blocking a site against the legal interest threatened or injured by the dissemination of the contents (see Rošnagel, Recht der Multimediadienste, Part 4, section l8, marginal no. 41).

As set out above, in my opinion the exclusion of domain in domain name servers (DNS) is sufficient and permissible in the present state of the art. This type of blocking can be carried out simply by configuring the DNS and incurs very low personnel costs on only one occasion. There are no material costs.

[...]

## 3.3 Follow-Up

Some Providers were appealing for the blocking orders but the district government rejected the appeals in July 2002. In the notification of the rejection the district government hinted that it preferred "DNS-Blocking" over other blocking techniques and stated that "this method can be implemented through easy configuration of the server and would only require a one time small amount of work and no materials."

Several providers filed complaints at local courts. This would have resulted in the blocking orders being frozen until the courts came to a decision. In September 2002 the district government used urgency orders ("Anordnung der sofortigen Vollziehung") to change that. To protect the public interest, it argued, the blocking orders should be in place until the courts decide. Several providers filed complaints again against this urgency order. Five local courts approved the urgency order and one rejected it. The court of appeal for the urgency orders then approved the urgency orders. Up to now no court has decided about the blocking orders.

After some hinting by the court of appeals the district government retracted not only the urgency order, but also the blocking order itself against the provider Telefnica. The reasoning was that Telefnica is not an access provider but an backbone provider and configuration of blocking in their carrier-class network would be to burdensome [Sch03].

## 3.4 What does the blocking order actually order?

It is very hard to understand what the blocking order actually wants to accomplish. There is at least certainty to some degree in the fact that the blocking order only talks about the Web. The whole blocking order uses exclusively vocabulary related to the Web and URLs referencing traditional web content reachable via the the HTTP protocol.

### 3.4.1 What is to be blocked?

Understanding which Web content is ordered to be blocked is much more difficult. While the blocking order's wording is clear that two Web pages referenced by URL - "which is the precise allocation criterium of the *individual* web page[s]",[3] - namely http://www.stormfront.org/ and http://www.nazi-lauck-nsdapao.com/ should be blocked, there is much indication that they actually want more than two Web pages to be blocked. In the reasoning the blocking orders first refer to the two "Internet pages", but later they speak about stormfront.org's "main page", which might be better translated by "home page" and use two other pages on that server and two pages only linked, hosted somewhere else, as reasoning for the blocking order. They also speak of the US "ISP" stormfront.org hosting right-wing content. The blocking order twice refers to stormfront.org's "Angebote" which was translated with "sites" but also can mean "offerings". Does the blocking order actually want everything hosted at the ISP stormfront.org to be blocked? Or are they aware of the fact that http://www.stormfront.org/ is similar to http://stormfront.org/ and would see both to be blocked? The blocking order asks for http://www.stormfront.org/ to be blocked but uses content referenced via the URL http://stormfront.org/german/zonen.htm which uses an different domain name.

For the "page" http://www.nazi-lauck-nsdapao.com the situation is similar: It is described as a "multi-paged home page" but also reasoned that "the homepage itself [contains many direct references]", then mentioned is a page with comments by Osama bin Laden which is located at http://www.nazi-lauck-nsdapao.com/binladen.htm and than again talk about the "whole offering" of nazi-lauck.nsdapao.com, but call nazi-lauck-nsdapao.com - contrary to stormfront.org - not an ISP, although both sell Web space, email accounts, etc.

The whole wording of the blocking orders is extremely confusing. They refer to "Internet pages" (Internet Seiten), "offerings" (Angebote), "main page" or "home page" (Hauptseite), "multi-paged home page" (mehrseitige Homepage), "Pages" (Seiten), "links" (Links), "Web pages" (Webseiten), "whole Website" or "whole offering" (Gesamtangebot), "references and links" (Verweise und Links) and "direct references" (unmittelbare Verweise) without explaining what these different terms should mean.

### 3.4.2 How should blocking be accomplished?

The problems understanding the blocking orders continue when trying to understand how blocking should be accomplished. In the first option for blocking there is very imprecise wording. DNS is - contrary to the claims in the blocking orders - not an acronym for "domain server" and not even for "domain name server". DNS stands for "domain name system" and describes a global, distributed database [AL99]. So no single provider "operates a DNS" as assumed by the blocking orders. While it can be safely assumed the blocking orders meant a recursive DNS server when they wrote DNS, the actual wording leaves the reader with a feeling that there are serious misconceptions about the working of DNS. This feeling manifests when reading the next sentence asking for queries to be "relayed to an invalid or an predefined page". There is no concept of pages in DNS, but only names and DNS resource records, so this simply can't be done. The only thing a recursive DNS resolver can return in this context are IP-addresses. However there is no such thing as an invalid IP-address or the IP-address of a predefined page. So the only action a recursive DNS server can take to vaguely resemble what this part of the blocking order might want to mandate, is returning the IP-address of a Web server which is sending a "predefined page" or redirecting the user to a "predefined page".

---
[3]Blocking order, emphasis by author.

Another question arises from the fact that the whole section explaining what should be blocked never explicitly mentions domains - so which domains actually should be blocked?

The wording of the whole passage and the fact that the most obvious way of DNS-tampering - refusing the query - is not allowed as a blocking measure and that the problem posed by different kinds of DNS resource records and handling of domain names in the same zone as the ones to be blocked are not mentioned at all, leaves many questions.

In the second blocking option the wording is again imprecise but it is possible to give somewhat consistent meaning to it. Still the blocking orders do not mention what can be considered "use of a proxy". Are the blocking orders fulfilled if a provider deploys filtering in a voluntary proxy as considered by [Ros03b] or do only enforced and interception proxies qualify? Also, which "individual web pages" or which URLs should be blocked?

The third blocking option mandates layer 3 IP-filtering. What this section doesn't state is which IP-addresses are to be filtered or at least how and how often to determine which IP-addresses to be filtered by resolving the domain names to be blocked. It was already shown that it is hard to find out which Web content the blocking order is ordering to be filtered - finding which IP-addresses belong to the web content is not that hard but the mapping from a domain name to an IP-address might change anytime, so it has to be monitored and updated [Ros03b]. As already mentioned the IP-address for stormfront.org changed 6 times in the last eight months.

The blocking orders are explicitly ordering layer 3 filtering, while layer 4 filtering is the obvious way of blocking access to Web content and has much fewer side-effects than layer 3 filtering. There is no reasoning in the blocking orders why layer 3 filtering is mandated.

Finally the blocking orders also state that at least DNS-tampering "can be carried out simply by configuring the DNS and incurs very low personnel costs on only one occasion". Supporters of the blocking orders like [Man02], but also authors opposing them like [Ros03a] even claim not only DNS-tampering but also filtering proxies and IP-filtering use only minimal resources at the provider. This claims can't be supported by facts. As shown above in many cases blocking via DNS-tampering can't be configured at all with popular software like BIND - unless you are accepting that you also harm a unknown number of domain names in the same domain. Blocking by tampering with IP-addresses in DNS A resource records also is extremely difficult to achieve without also affecting other DNS resource records. To do so constant maintenance and updating is needed. Also sound engineering practices - which are crucial to run critical infrastructure at carrier class level - dictate that *no* configuration change to a vital production system is trivial. Especially making software explicit behaviour it was not designed to do requires rigid change control including extensive testing and monitoring. In the legal literature only [Zim99] considers this fact, but [Sch03] suggests, that the courts also understand this - at least in some cases.

What is completely missing from the blocking orders is a description of what actual content is to be blocked. Web content is very volatile: Web servers get reorganized, domains get new owners. This was prominently demonstrated in the context of the blocking orders by the Website at www.front14.org: at fall 2001 this domain contained a right wing extremist portal but by spring 2002 there was a Web catalog at the same Address with no obvious political agenda. This underlines the need for identifying pages to be blocked not only by their location via an URL but also by their actual content. A machine readable description, e.g. in form of a cryptographic fingerprint could be easily generated in an automatic fashion and also could be also used for automatically determining if the content to be blocked still is residing at the URL which was blocked.

It is also noteworthy that the blocking orders set no emphasis on the users at all. The user experience of the blocking measures seems to be completely out of concern. One would have expected that the orders were crafted in a way that users trying to access blocked pages do not end up with obscure errors but get a clear message that the access to the content they have tried to reach was prohibited by the district government. While a democratic society actually has the right to suppress certain speech in certain circumstances, the fact that something has been suppressed must be clearly stated, instead of being hidden in obscure error messages. Suppressing unwanted speech clandestine seems not worthy of a democracy.

## 4 Current DNS-Situation

It seems that tampering with DNS is the preferred way by providers and the district government of hampering access to certain Websites. While researching the blocking orders I was only able to verify IP-filtering at one Institution with additional rumors about IP-filtering at a second institution.

As shown above there are several implementation choices for DNS-tampering which have very different implications for affecting other services, privacy of users and user experience. To assess the current state of DNS-tampering empirical research was conducted by doing a DNS survey on recursive DNS servers at providers in NRW.

### 4.1 Survey Methodology

The goal of the survey was to get an overview of the current implementations of DNS-tampering. Most accurate results would have been obtained by signing up with all providers in NRW, accessing the Internet via their dial-in services and then querying the providers recursive DNS servers as they where set via the dial-in process, for the domains to be blocked. This method faces several serious drawbacks: Not only the costs and workload to carry through sign-up process with dozens of providers but also the fact that some providers in the cable and xDSL sector are not accessible via dial-up and others like universities make their services available only to closed groups of subscribers. Another problem is the fact that there is no authoritative list of access providers in NRW available. Even a list of providers which received the blocking orders could not be obtained.

To overcome this a sample of name servers at providers in NRW was generated semi-automatically and queried over the Internet. First a list of domain names of providers in NRW was obtained by querying the provider section of http://web.de/, a big German Internet catalog, for providers with ZIPs mapping to NRW. In addition a list of NRW's Universities' and polytechnic Universities' domain names was obtained from the Web. These domains where queried for canonical host names often used by recursive servers. In addition zone transfers for the domains where tried, to find subdomains for which the procedure was repeated in a recursive fashion. Additional recursive DNS servers where found by querying the DNS NS resource records for the domains obtained by the steps above. For some high-profile providers where no recursive DNS servers were found by the above procedure a Web search on provider name and "DNS Server" was carried out and results added to the list. Finally the list was purged of name servers obviously not belonging to providers in NRW. Some providers from different German states where kept in a separate list as a control group.

Using this list of recursive DNS servers for a survey has certain disadvantages. First there is no guarantee that the users of a provider use the same recursive DNS servers than the ones which were accessed by the survey. Also the servers may refuse queries from the Internet or deliver different data to external users than to internal ones. Keeping these constraints in mind, the survey methodology still seems to be an appropriate tool for collecting an overview of DNS-tampering in the wild. But one must carefully keep in mind that the data was not gathered in a way guaranteeing it to be representative. Specifically drawing conclusions on the percentage of providers deploying blocking is nearly impossible.

Between the 10. and 13. of May 2003 the recursive DNS servers on the list were queried for records of type A, ANY, MX, NS and SOA for several names within the nazi-lauck-nsdapao.org, stormfront.org front14.org and rotten.com domains. Data for some other domains was requested as a control group.

DNS-replies varied far more than expected. This made an automatic categorisation difficult. Only DNS MX and A resource records for stormfront.org, www.stormfront.org, kids.stormfront.org, rotten.com and md.hudora.de where analyzed. Automatic methods were used for screening the resulting data and singling out replies which indicated tampering. Reference data was obtained form a recursive DNS server known to be untampered. Replies from other recursive DNS servers were compared to this data. Replies containing mostly the same header fields as the reference data were considered untampered.[4] Comparing the answer section of the replies turned out that many recursive DNS servers are modifying the preference parameter of DNS MX resource records returned. Also the TTLs of replies varied. Replies only varying in the TTL or MX-preference where considered untampered. Replies with a REFUSED status where saved for further analysis in the future, replies with a NXDOMAIN status where logged as being tampered and replies with an NOERROR status, but failing the comparison against the reference data where dumped in a format resembling output of the UNIX tool "dig" for further analysis by human inspection. Then replies per recursive DNS server where aggregated to one group per provider by querying WHOIS records for the recursive DNS server's IP-address. Where this data

---

[4]The following header fields had to have the same values to consider two headers similar: ancount, opcode, qr, rcode, rd, status, tc. Oddly there where a substantial number of replies which had the aa flag set. Reasons for this need further investigation.

seemed inappropriate, additional research on the Web was carried out.

## 4.2 Findings

Datasets returned by recursive DNS servers of different providers varied greatly and were hard to group together in an intuitive fashion. We will first look at results grouped by what they actually block and in which way they do this. Following that, providers with strong distinctions will be shown. Finally data will be summarized. All results will refer only to domain names in the stormfront.org domain unless there are notable exceptions with other domain names.

### 4.2.1 Results and References / Examples

**hijacking / NXDOMAIN combination**  Some providers manipulate their recursive DNS server in a way that queries for www.stormfront.org result in some kind of answer which should accomplish blocking. Results to all other queries in the domain stormfront.org result in NXDOMAIN replies stating that the queried name does not exist. Interestingly queries for requests to the stormfront.org domain name itself result not in an NXDOMAIN reply but in an empty answer. Both types of reply result in domain names as kids.stormfront.org and stormfront.org and also mail service being unavailable. Implementation details vary widely:

- *Kamp Netzwerkdienste GmbH* returns the address of an machine on their network for www.stormfront.org. The Web server on the Kamp machine redirects to a non existent page on a server of the district government.[5] This results in a "file not found" error to the user.
- The *Ruhr Universität Bochum* and *Evangelische Fachhochschule RWL* return a so called "CNAME" for DNS A resource record queries to www.stormfront.org which points to the address of an machine at the Ruhr Universität Bochum. This machine redirects requests for http://www.stormfront.org/ to a page[6] claiming in German language: "The content of the requested page is deemed illegal by German law".
- *tro:net GmbH* returns the IP-address of one of tro.net's machines to queries for DNS A resource records for the domain name www.stormfront.org. The user accessing http://www.stormfront.org/ is shown a page claiming in German language about www.stormfront.org that "this domain is already registered, a Website is being built".
- *tops.net GmbH & Co. KG* returns for IP-address queries to www.stormfront.org a DNS CNAME resource record pointing to sperrungsverfuegung.tops.net. For stormfront.org the IP-address of sperrungsverfuegung.tops.net, is returned directly. This machine returns a text with a very complete explanation of the blocking situation. It also tries to store a so called cookie on the user's Browser enabling tops.net to re-identify it should the user access other Web pages at tops.net.

---

[5]Namely http://www.bezreg-duesseldorf.nrw.de/cat/SilverStream/Pages/presseframe?BeitragsID=7008.
[6]Namely http://www.ruhr-uni-bochum.de/www-rz/zollehcc/_rz/unzulaessig.htm.

**NXDOMAIN** *Global Village GmbH*, *INCAS AG* and *NetCologne GmbH* all reply for any query in the stormfront.org zone that the requested data does not exist. For requests to the stormfront.org domain name themselves they return an empty reply. Both result in "Host not found" error messages for every Web request or email attempt.

**www only hijacking** Some Providers leave the domain stormfront.org itself untouched and tamper with the www.stormfront.org domain name only. This results in Web access to http://stormfront.org/ and email to any@stormfront.org being unaffected, while DNS MX resource records for mailing to any@www.stormfront.org are damaged in a way that email to these addresses would probably bounce with an error message.

- *Bogs Marketing & IT GmbH*, *LB MikroComputerTechnik*, *pixelhouse media services* and *UUnet* return to users querying for www.stormfront.org the IP-address of a machine at UUnet which deploys a Web server redirecting users trying to access http://www.stormfront.org to a page at eco.de,[7] informing about the situation.

- *Westend GmbH* is returning the IP-address of one of their own machines running a Web server redirecting the user to the eco page informing about blocking mentioned above.

- *Xenologics Networks & Communications GmbH* returns the IP-address of a machine on their own network for queries to DNS A resource records, redirecting again to eco. It seems this company has tried to avoid blocking email: they did so by besides inserting DNS A resource records pointing to their own machine also adding DNS MX resource records in the www.stormfront.org zone file. But the data returned for MX queries by Xenologics is not consistent with the data returned by the authorative nameservers for stormfront.org. It can be safely assumed that Xenologics put once correct DNS MX resource record data in the zone file but failed to update it regularly. So the actual configuration still results in problems sending email to addresses at www.stormfront.org.

- *ST-oneline InterNet Service Provider GmbH*, *Versatel Deutschland GmbH und Co KG*, *Pironet NDH AG* and *wel.de Gesellschaft für Informationsdienste* return the address of a machine at Pironet NDH AG, redirecting to a page by eco[8] which is advertising a completely unrelated product. Some of ST-oneline's, Versatel's and wel.de's recursive DNS servers return untampered data.

- *Ginko / QSC AG* do not redirect users which try to access http://www.stormfront.org/ to a special page. Instead they return an empty zone for the name www.stormfront.org resulting in "Host not found" error messages to the user.

---

[7]Namely http://www.eco.de/servlet/PB/menu/1188401_l1/index.html.
[8]Namely http://www.eco.de/greenspot.

- *regioconnect GmbH* deploys name astrayment by answering queries for the address of www.stormfront.org with IP-address of 127.0.0.1. This results in the wide array of possible disturbing results outlined above.

|  | should be blocked | accessible | blocked | obscure error | error rate |
|---|---|---|---|---|---|
| stormfront.org | yes | 12 | 4 | 11 | 44% |
| www.stormfront.org | yes | 0 | 12 | 15 | 0% |
| kids.stormfront.org | no | 12 | 0 | 15 | 56% |
| rotten.com | no | 24 | 3 | 0 | 11% |

Table 1: Effectiveness of blocking Web pages via DNS-tampering

Some providers do manipulations which do not fit in the categories above.

- *Universität Essen* is doing name astrayment by returning the IP-address 1.1.1.1. for address queries to www.stormfront.org and stormfront.org. This IP-address is not controlled by the University of Essen, but IANA. At the moment IANA has announced no routes for this IP-address to the Internet. This usually results in data designated to 1.1.1.1 wandering around the Internet for some time until finally the data is discarded and the sender is informed that 1.1.1.1 is not reachable. This will result in "Could not connect" error messages to the user after some time.

  It seems the University of Essen tried to use a "invalid" IP-address. But there are no invalid IP-addresses. The university of Essen has no right to direct traffic to IP-addresses not allocated to it if the "owner" of the IP-address does not consent. I strongly doubt that the IANA has consented in using 1.1.1.1 to do DNS-tampering. At least one name server inside the University does return original, untampered data.

- *ISIS Multimedia Net GmbH* is doing name astrayment by returning the IP-address 127.0.0.1 to all DNS A resource record queries within the stormfront.org domain. Queries to all other kinds of DNS resource records result in NXDOMAIN or empty replies. This means as shown above that users trying to access Web pages on any domain name in the stormfront.org domain see their own computers' content or their providers Web pages or get an error message.

- *Fachochschule Münster* a polytechnic University and *Citykom Münster GmbH Telekom-munikationsservice*, both located in the city of Münster, do some very active manipulations of DNS data. For www.stormfront.org FH-Münster returns the Address 127.0.0.1 which is the address of the users own computer. See above for results of this manipulations. Citykom returns the address of a server at Citycom which sends invalid data to the requesting browser breaking the HTTP specification and leaving the user with an error message. It can be assumed that Citykom wanted to redirect users to a page at the district government's server[9] but made an error when configuring this behaviour. If queried for any other DNS A resource records in the stormfront.org zone, both providers reply with NXDOMAIN or empty answers.

---
[9]Namely http://www.bezreg-duesseldorf.nrw.de/cat/SilverStream/Pages/presseframe?BeitragsID=7008.

The remarkable modification is that of DNS MX resource records, which direct where email is sent to: both organization's recursive DNS servers explicitly state when queried for DNS MX resource records that email to stormfront.org should be delivered to machines in their own organization. Citycom is also hijacking mail to www.stormfront.org in the same manner.

- *BergNet Onlinedienst GmbH*, *Oberberg Online Informationssyseme GmbH* and *Vision Consulting Deutschland OHG* which seem to share some infrastructure, all return to DNS A resource record queries on www.stormfront.org and stormfront.org the IP-address of a machine at *Vision Consulting* which is presenting users a German language page titled "Routing denied: The page you wish for will not be shown." followed by some paraphrased German laws about illegal Internet content. All other queries in the domain stormfront.org result in NXDOMAIN or empty replies.

  Also requests to http://www.rotten.com/, http://rotten.com/, http://www.front14.org/ and http://front14.org/ get redirected to the page mentioned above. All other requests in the domains rotten.com and front14.org result in NXDOMAIN or empty replies. One of Vision Consulting's nameservers returns untampered DNS data.

|  | should be blocked | unharmed | broken | error rate |
|---|---|---|---|---|
| postmaster@stormfront.org | no | 11 | 16 | 59% |
| postmaster@www.stormfront.org | no | 0 | 27 | 100% |
| postmaster@kids.stormfront.org | no | 12 | 16 | 59% |
| postmaster@rotten.com | no | 24 | 3 | 11% |

Table 2: Side-effects on email of blocking Web pages via DNS-tampering

### 4.2.2 Summary

- Many Providers seem to assume that http://stormfront.org/ is not to be blocked. As shown in table 1 of 27 providers 12 didn't block it at all and 11 possibly blocked it only by accident. This is an error rate of at least 44%.

- Keeping email usable seems to be no issue to most providers. As shown in table 2 all providers block at least some email. A single provider has tried to reduce email blocking by not tampering with DNS MX resource records, but failed in this effort. All other seemingly didn't even try to keep email from being affected.

- Privacy of users trying to access the blocked pages seems to be no issue to most providers. One provider is even using - possibly by accident - cookies, two providers reroute email to their own systems, 10 providers return DNS A resource records at machines located at other providers, 12 providers allow third parties to monitor redirects leading to them, where in two cases the third party is the district government itself.

- Informing users of what actually is happening seems of no priority. Web accesses to blocked content results at 11 providers always in confusing errors and at all other providers at least in some cases in confusing errors.

- Configuration of DNS-tampering seems to be difficult. At least 30% of the providers have created major misconfigurations besides being overrestrictive or underprotective.

- Sites not directly mentioned in the blocking order and run by different persons than the sites which were mandated to be blocked where substantially hit by erroneous blocking. http://kids.stormfront.org/ is blocked by 58% of the surveyed providers. http://www.rotten.com/, which the district government in 2001 briefly considered to be blocked, is blocked by 11% of the providers.

- Compliance with the blocking orders seems to be next to impossible. As shown in table 3, even when stretching the legal principles to the maximum and interpreting the blocking orders in the broadest possible way, only 55% of the providers comply with them. Interpreting the blocking orders more reasonable in a way that they try to protect non-Web communication from being blocked, we see no single provider complying. With this interpretation 45% underprotective and overrestrictive at the same time while the remaining 55% are "only" overrestrictive.

|   | underprotective | complying | overrestrictive | correct |
|---|---|---|---|---|
| 1 | 0 | 0 | 27 | 0% |
| 2 | 12 | 0 | 27 | 0% |
| 3 | 12 | 0 | 27 | 0% |
| 4 | 0 | 12 | 15 | 45% |
| 5 | 12 | 0 | 15 | 0% |
| 6 | 12 | 15 | 0 | 55 % |

1. block Web access on www.stomfront.org and leave other services untouched where possible
2. dito but also block stormfront.org
3. dito but block Web access to any name in the domain stormfront.org
4. block all communication to www.stomfront.org
5. block all communication to stormfront.org and www.stomfront.org
6. block all communication to any name in the domain stormfront.org

Table 3: Possible interpretations of the blocking order and the state of provider compliance

## 5   Conclusion

While it should certainly not be argued that government mandated blocking of foreign Web content is mainly a technical problem, not looking at the technical aspects misses important factors. It was shown that blocking from a technical point of view has considerable trade-offs. DNS-tampering theoretically can't be done without hurting an unknown amount of other content and services. Even if this is accepted, configuring DNS-tampering with minimal side-effects is very difficult with DNS software in wide use today. Forcing users to use HTTP proxies opens their Web experience to a wide array of manipulation and logging which has to be carefully considered. Many circumvention techniques are today only a theoretical possibility or will be used only by a small fraction of Internet users. The blocking orders themselves are technically imprecise to a degree that they are nearly incomprehensible. This can be proofed by the fact that in any possible interpretation nearly half of the surveyed providers do not comply with the blocking orders. The blocking orders also fail to discuss while less intrusive measures like layer 4 IP-filtering are not preferred over layer 3 IP-filtering.

Surveying the providers implementing DNS-tampering indicates that many are underprotective and still more are overrestrictive. Being underprotective and overrestrictive seems to be a common pattern in suppressing Web content [Hun00] [NW99].

Further research on blocking is needed. Foremost methods for estimating the side-effects of blocking need to be developed since this estimates are crucial for the government making informed decisions about blocking. In addition a common language for precisely defining the content to be blocked seems to be badly needs. Also the actual effects of the blocking orders should be monitored - not only from a technical point of view but also from a criminological perspective.